\documentclass[submission,copyright,creativecommons]{eptcs}
\providecommand{\event}{QPL 2011} 
\usepackage{breakurl}             
\usepackage{graphicx}
\title{A Framework for Heterotic Computing}
\author{%
Susan Stepney
\institute{Department of Computer Science,\\University of York, UK}
\email{\quad susan@cs.york.ac.uk}
\and
Viv Kendon
\institute{School of Physics and Astronomy,\\University of Leeds, UK }
\email{\quad V.Kendon@leeds.ac.uk}
\and
Peter Hines
\institute{Department of Computer Science,\\University of York, UK}
\email{\quad phines@cs.york.ac.uk}
\and
Angelika Sebald
\institute{Department of Chemistry,\\University of York, UK}
\email{\quad angelika.sebald@york.ac.uk}
}
\def\titlerunning{A Framework for Heterotic Computing}
\def\authorrunning{S. Stepney, V.Kendon, P.Hines \& A. Sebald}

\begin{document}

\maketitle

\begin{abstract}
Computational devices combining two or more different parts, one
controlling the operation of the other, for example, derive their
power from the interaction, in addition to the capabilities of the parts.
Non-classical computation has tended to consider only single
computational models: neural, analog, quantum, chemical, biological,
neglecting to account for the contribution from the experimental controls.
In this position paper, we propose a framework suitable for
analysing combined computational models,
from abstract theory to practical programming tools.
Focusing on the simplest example of one system controlled by another
through a sequence of operations in which only one system is
active at a time, the output from one system becomes the input
to the other for the next step, and \textit{vice versa}.
We outline the categorical machinery required
for handling diverse computational systems in such combinations,
with their interactions explicitly accounted for.
Drawing on prior work in refinement and retrenchment, we suggest an
appropriate framework for developing programming tools from
the categorical framework.  We place this work in the context of
two contrasting concepts of ``efficiency'':  theoretical comparisons
to determine the relative computational power do not always reflect
the practical comparison of real resources for a finite-sized
computational task, especially when the inputs include (approximations
of) real numbers.  Finally we outline the limitations of our simple
model, and identify some of the extensions that will be required to treat
more complex interacting computational systems.
\end{abstract}

\section{Introduction}
\label{sec:intro}

Classical computation theory is epitomised by the Turing machine paradigm. 
We are concerned with more
diverse models of computation, in particular determined by the physical
properties of the system used as a computer \cite{SS-PhysicaD-08}. 
A broad range of experiments and theory is being developed to
investigate the computational capabilities of
chemical \cite{Kuhnert_1989,Motoike2005107,Toth_1995},
biological \cite{Adamatzky_2007,Amos},
quantum \cite{spiller05a},
optical \cite{Woods2008,Tucker2010},
and
various analog \cite{lloyd99b,silvagraca04a,mills08a} computational substrates.
Given that we have different types of computational devices,
not necessarily Turing universal, it is natural to ask
how to {\sl compose} them, and to ask about the computational
power of the composition. 
We term such composed systems
\textsl{heterotic computers}\footnote{%
\textsl{Heterotic}, from the
Greek \textsl{heterosis}, a term in genetics meaning ``hybrid vigour''.
}.

The computational power of a given physical system is determined 
not only by the operations available to manipulate the system,
but also by the type of data that can be encoded in the system
and the measurements available to decode the result of the computation.
When composing different systems, information must pass between
them, making these data types and measurements relevant throughout
the computation.  
This is in contrast to classical complexity analysis,
which focuses on the operations that perform the computation, and
assumes that data input and output are trivial in comparison.
More care is generally taken when using non-standard computational models.
For example, in quantum computing, DiVincenzo's checklist \cite{DiV00}
first identifies a physical system that can represent a qubit, then
identifies a set of operations sufficiently rich to provide universal
quantum computation. 
Output from quantum systems is also non-trivial, since measurements
cannot determine the full quantum state with certainty.  Extra procedures
in the algorithm are required to ensure the measurement gives a useful
output with high probability.
However, this analysis still focuses on the quantum processor without
giving explicit account of the role of the classical control systems.

Thus, we need a framework that not only allows different models of
computation to be compared and contrasted, but also allows us to compose
different models and determine the resulting computational power,
as motivated in \cite{KSSBHW11}.
In this position paper, we provide more details of the categorical
tools required to accomplish this.
Together with a refinement/retrenchment approach to support program
development, these would provide the tools to 
determine the combined computational power of the heterotic computer.
The paper is organized as follows:
In \S\ref{sec:het} we summarise prior work on several heterotic
systems: measurement-based quantum computing; NMR classical computing;
qubus quantum computing.
In \S\ref{sec:fwk} we outline the categorical framework, in the context of
a simple two-layer computational architecture,
and outline a semantic basis and refinement approach.
In \S\ref{sec:ref} we describe how to create the programming tools from
this framework, using a modified refinement based method. 
In \S\ref{sec:conc} we summarise and outline the next steps for this work.

\section{Heterotic computational systems}
\label{sec:het}

\begin{figure}[tb!]
\centerline{
(a)~~\scalebox{0.2}{\includegraphics{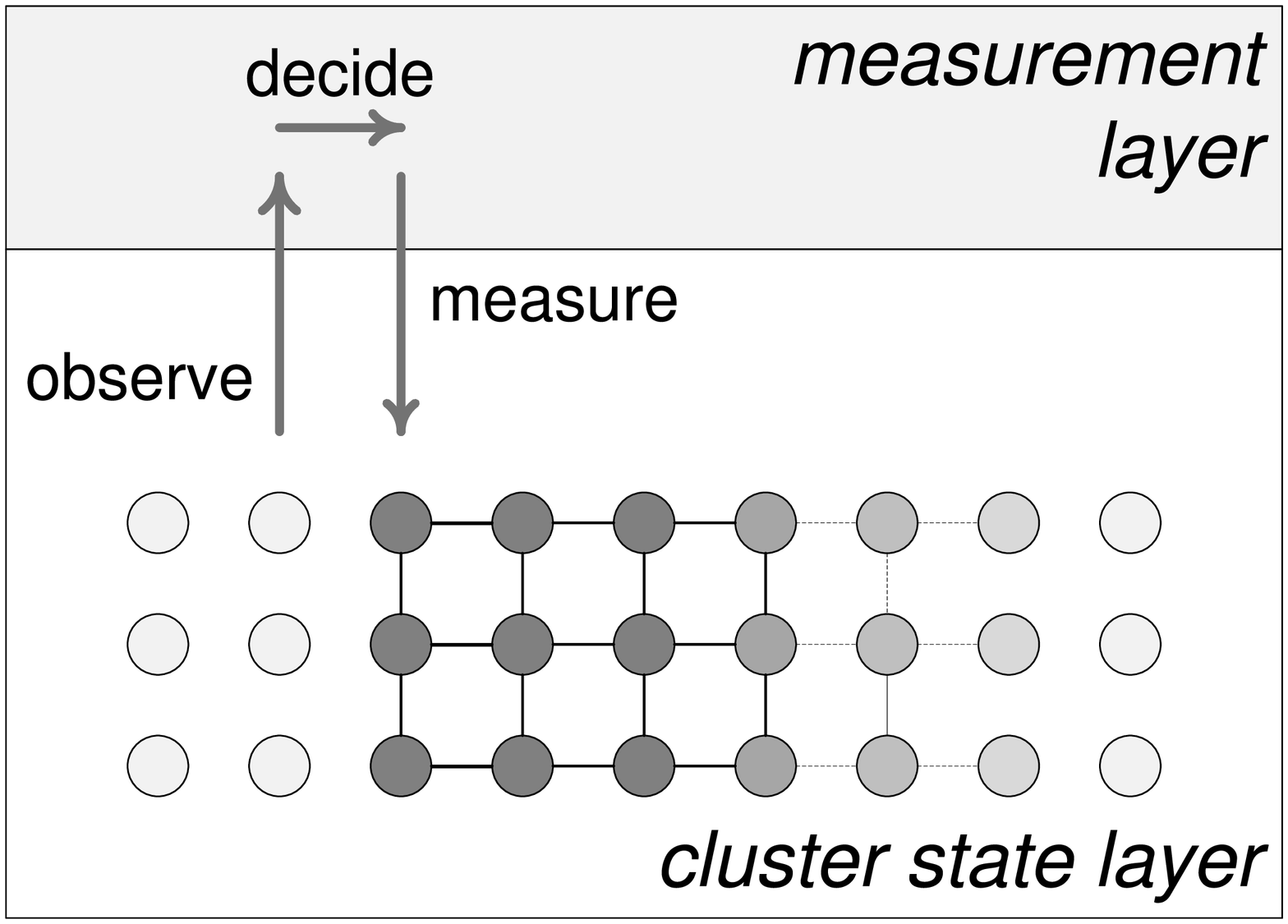}}
~~~~(b)~~\scalebox{0.2}{\includegraphics{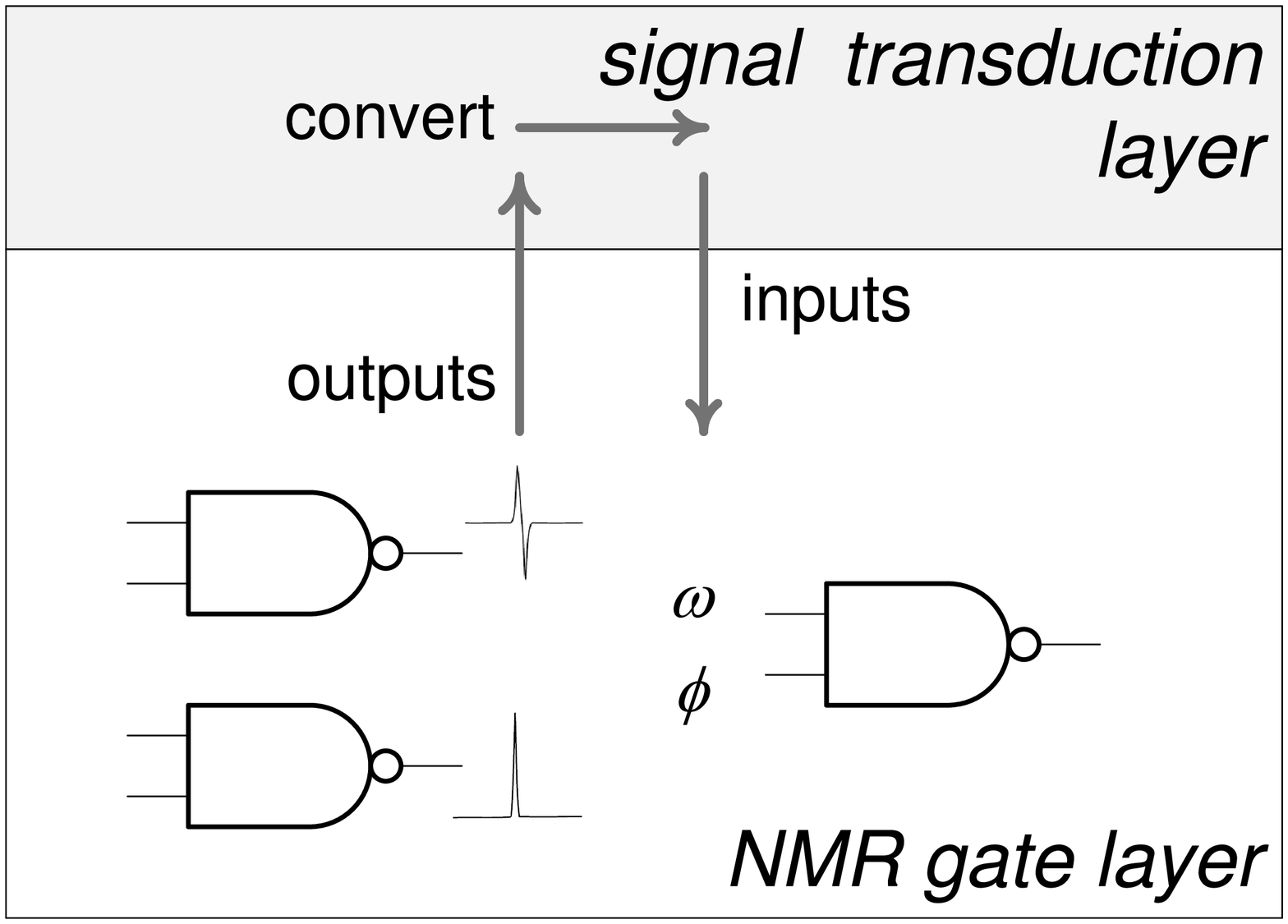}}
}
\vspace{1em}
\centerline{
(c)~~\scalebox{0.3}{\includegraphics{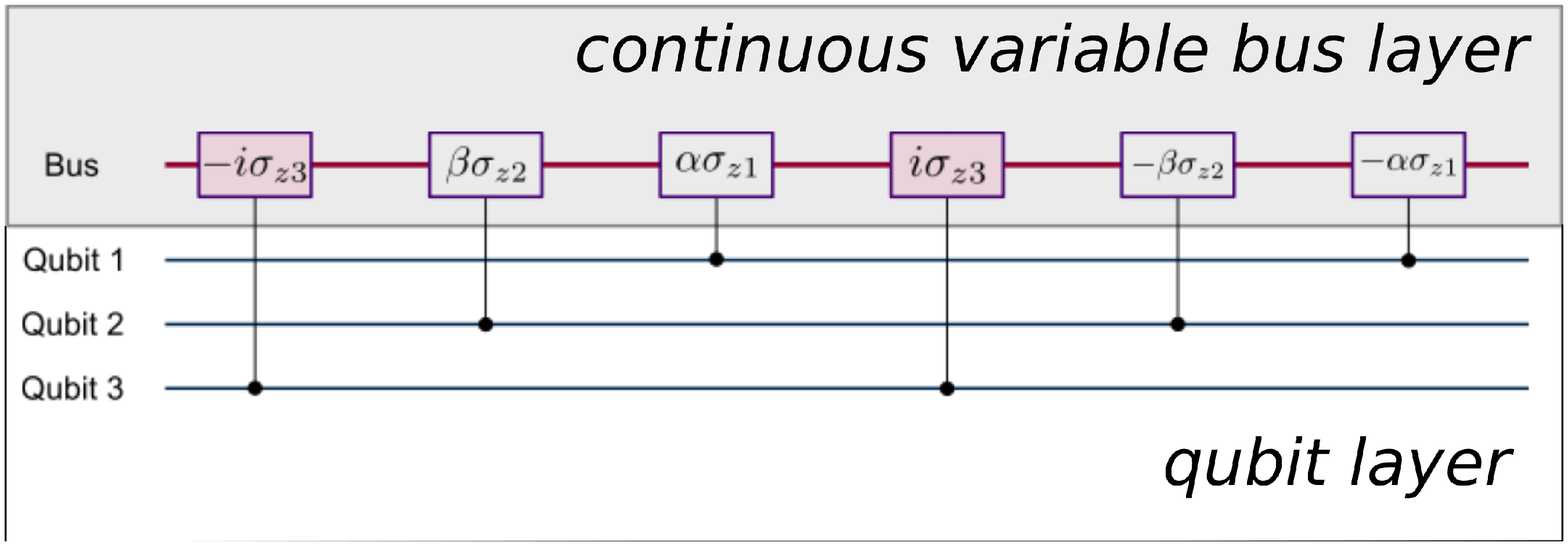}}
}
\caption{
(a) Measurement-based quantum computer.
The base layer is a cluster state. 
The control layer performs measurements on the base layer,
thereby changing its state;  the control layer uses the observed results 
of a measurement to decide what measurement to perform next.  
(b) Classical NMR computer \cite{RBSS10}. 
The base layer gates are implemented as NMR experiments: inputs are frequencies $\omega$ and phase delays $\phi$; outputs are the integrated output signal. 
The control layer performs ``signal transduction'': taking the integrated output,
interpreting it as a 0 or 1, and converting that to the appropriate
physical input signal.  
(c) Qubus quantum computer.
The base layer is qubits, the control layer is a coherent state,
which can interact with several qubits at the same time, enacting
the gates between the qubits.  There are no measurements in this fully
quantum example, the qubit state determines the interaction with the bus,
which in turn changes the qubit state according to the externally chosen
order in which it interacts with the qubits.
}
\label{fig:cluster}
\end{figure}

The role of the classical controlling system in quantum computation
was first noted by Josza \cite{Jozsa05},
while demonstrating the equivalence of measurement-based
and teleportation-based quantum computing schemes.
In measurement-based quantum computing (MBQC), also known as cluster state,
and as one-way, quantum computing \cite{RB01}, an entangled resource
of many qubits is prepared, then the computation proceeds by
measuring the qubits in turn.  The outcomes from the measurements
feed forward to determine the type of measurement performed on
the next qubits (figure \ref{fig:cluster}a).
It was not until 2009 that
Anders and  Browne \cite{AB09} realised that the classical
computation required to control and feed forward information in MBQC
is a crucial part of the computational power.
Applying measurements without feed-forward is efficiently classically
simulable, as is (trivially) the classical part of the computation. 
However, the combination of the two is equivalent to the quantum circuit model, 
which is not (efficiently) classically simulable.
Thus the combination of two or more systems, to form a new
computational system composed of several layers, can be in
a more powerful computational class than the layers acting separately.

Equivalent examples have been described in the realm of
classical unconventional computation.
In experiments using liquid state NMR to perform simple
gate logic \cite{RBSS10} such as NAND, the instruments controlling the NMR
pass the outputs of one gate through to the inputs of the next
(figure \ref{fig:cluster}b).
As with MBQC, these controls play an essential role in the computation,
but by themselves do not perform the gate logic.
Using NMR to do classical computing involves choosing a subset of the
available parameters suitable for representing classical bits, and
restricting the operations to keep the spin ensemble in a fully determined 
classical states.  In this way, more robust operations are obtained
at the expense of not exploiting the full capabilities of the physical 
system.  
Prior work on computation using NMR mostly deals with implementations of 
quantum computations, predominantly based on solution-state NMR
experiments \cite{Jones2011},
with some examples exploiting solid-state NMR \cite{Cory2000}.
As a step towards characterizing the computational power of
the NMR system, Bechmann et al \cite{BSS11}
have produced a preliminary classification of the experimental
NMR parameters for implementing classical logic gates.
This work has been extended to take advantage of the inherently
continuous nature of the NMR parameter space of non-coupled spin
species \cite{SS-ICES2010} by implementing continuous gates,
so the combined system performs an analog computation. 
However, the extent to which the control layer contributes to the computational 
power of quantum or classical NMR computing has yet to be analysed.

The theory of ancilla-based quantum computation \cite{AOKBA10} has
been abstracted and developed from MBQC, into a framework
where a quantum system (ancilla) controls another quantum system (the
qubits), with or without measurement of the ancilla system during 
the computation.  This framework is capable of modelling many types of
hybrid quantum computing architectures.
When the role of the ancilla system is played by a continuous variable
quantum system instead of a qubit or qudit ($d$-dimensional quantum system)
further efficiencies become available.  The qubus quantum computer
uses a coherent state as the bus, which has two quadratures, which
act as two coupled continuous variable quantum systems.
This type of ancilla can interact with many qubits at the same time,
allowing savings in the number of basic operations required for gate
operations \cite{BDKM10} and for building cluster states
\cite{HBMK11,BHKM11}.  
Figure \ref{fig:cluster}c shows a sequence of six operations 
that performs four gates, one between each possible pair of the three qubits.
Each gate performed separately would require two operations,
thus this sequence saves at least two operations over standard methods,
more if the qubits have to be swapped to adjacent positions for direct gates.
Typically, this provides polynomial reductions in
the number of elementary operations required for a computation, when
compared with interacting the qubits directly.

\section{Towards a categorical heterotic framework}
\label{sec:fwk}

\begin{figure}[tp] 
\centering
\scalebox{0.65}{\includegraphics{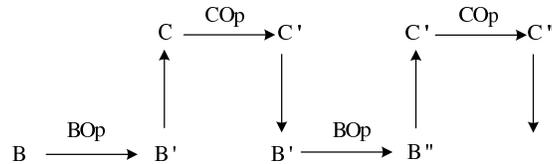}}
\caption{The stepwise interactions between a base computation (state $B$, state change $BOp$) and a controller computation (state $C$, state change $COp$): 
the input to one is the output from the other.}
\label{fig:hetero}
\end{figure}
The examples in \S\ref{sec:het} can all been depicted with the same
structure of a base layer and a control layer (figure~\ref{fig:hetero}).
This can be generalised to multiple layers, each controlling the
layer below, and being controlled by the one above, and 
to layers with feedback loops that couple non-adjacent layers.
Here we focus on the simplest heterotic model,
in which just two computers are coupled, one controlling the other.
For now, we take as given the particular division into layers:
we do not need this to be a unique decomposition in what follows. 

The pattern of computation and communication alternates between the
two layers (figure~\ref{fig:hetero}). 
In this basic model, the state of one layer does not change during the computation by the other (for example, the control layer remains in state $C'$ as the base layer evolves from $B'$ to $B''$).
The basic model allows a physical implementation where the state continues
to evolve, if its subsequent computation depends only on its input
(either it is essentially ``reset'' to the previous state, 
or the input fully determines what happens next).
This case holds for our motivating examples in figures~\ref{fig:cluster}a
and~\ref{fig:cluster}b, although they have not yet been explicitly
cast in the framework.  In the qubus example, figure~\ref{fig:cluster}c,
the layers shown evolve only while interacting with each other.
However, single qubit gates applied directly to the qubits can be inserted
whenever the qubus is not connected to the qubit in question, which would
then be an example of two separate controlling layers doing different tasks.
Furthermore, the coherent state (a quantum state)
acting as the bus itself has a classical control layer (not depicted), which  determines the parameters ($\alpha,\beta$)
in the interactions with the qubits.
This architecture thus goes beyond our simple starting point of two
coupled computers, and serves to remind us that extensions to the basic
model will be required.

One of the goals is a form of \textsl{refinement calculus
for heterotic computers}, suitable for use by the working programmer,
to enable the full power of such systems to be exploited.
However, producing such a framework first requires theoretical input. In
particular, we need a suitable form of semantics on which the refinement
calculus is based.
Such models exist for individual systems, for example,
classical analog computation has been modelled in several ways, from the
traditional approaches based on differential equations, to interval-based
analyses relying on domain theory.
Classical probabilistic
computation can be modelled via categories of stochastic relations, and
non-determinism frequently requires categories of relations, or
constructions based on the power set functor.
For heterotic computing, the theoretical challenge is to give a formal
description of how such systems may interact in non-trivial ways.
Due to the wide range of heterotic computing systems under consideration,
we aim for an abstract categorical semantics,
and seek concrete instantiations where appropriate.

Given two
dissimilar systems $A$ and $B$, and models of each of these in distinct categories $\mathcal{C}_A$ and $\mathcal{C}_B$, we require a formal setting in
which both  the joint system, and the non-trivial interactions between systems $A$ and $B$, may be modelled. 
If we wish to model a joint system \emph{without} considering interactions, the product category $\mathcal{C}_A \times \mathcal{C}_B$ is the natural choice; however, for our purposes, it is entirely inappropriate. The real object of study (and, we claim, source of computational power) is found in the non-trivial interactions between the subsystems.

How, then should we describe interactions between computing devices whose models are to be found in distinct categories? One approach 
would be to find some larger encompassing category, sufficiently broad and general to model both devices (similar to the way that both classical probabilities and complex phases may be combined in the density matrix formalism of quantum mechanics). However, the downside of this approach is that, with highly dissimilar devices, the required framework must be excessively abstract or general. There is also the more philosophical objection that this approach would be trying to treat our interacting systems as a single system in some more general setting, missing the motivation of studying the \emph{interaction} of distinct systems for its source of computing power. 

So instead, to model interactions between systems $A$ and $B$, we rely on some structure-preserving map from models of system $A$ to models of system $B$, and vice versa. These must be functors $\Gamma : \mathcal{C}_A \rightarrow \mathcal{C}_B$ and $\Delta : \mathcal{C}_B\rightarrow \mathcal{C}_A$. The question is, what further categorical properties must these be expected to display? 

As a motivating example, we consider categorical structures that are at the core of many computing systems, and consider how they can be either generalised or relaxed, in order to deal with systems based on interacting distinct systems. In categorical models of logic and computation,  the notion of a closed category -- usually monoidal closed -- is often fundamental. In logical systems, monoidal closure provides the structure necessary to model  cut-elimination, and given a computational interpretation of logical systems (commonly via the Curry-Howard isomorphism) this interprets as 
 $\beta$-reduction in lambda calculus \cite{LS}. Other logical or computational interpretations are available, from compositionality in models of Turing machines \cite{PH2}, to the essential categorical structure of teleportation in quantum computation \cite{AC04}.

A monoidal category $\mathcal C$, has a functor (the 
{\em monoidal tensor}) $\otimes: {\mathcal C}\times{\mathcal C}\rightarrow {\mathcal C}$,
satisfying 
$(A \otimes B) \otimes C \cong A \otimes (B \otimes C)$ 
together with a unit object $I$ satisfying $A \otimes I \cong A \cong I \otimes A$. (The families of arrows exhibiting these isomorphisms must satisfy additional {\em coherence} and {\em naturality} conditions; see \cite{MCL} for more details).
A monoidal category is {\em monoidal closed} when there also exists a functor (the 
{\em internal hom}) $[\underline{\ \ } \rightarrow \underline{\ \ }]: {\mathcal C}^{op}\times {\mathcal C}\rightarrow {\mathcal C}$ that satisfies 
\begin{equation}
 {\mathcal C}(A\otimes B,C) \ \cong \ {\mathcal C}(B,[A\rightarrow C]) 
 \end{equation}
This is a canonical example of an adjunction. Further, in the very special case where the system is {\em untyped} (so all objects of $\mathcal C$, excluding the unit object,  are isomorphic), we recover the familiar untyped equations
$D \ \cong \ D \otimes D \ \cong \ [D\rightarrow D] 
$ 
providing models of {\em universal computation} (e.g. the C-monoids of \cite{LS} or the untyped compact closure of \cite{PH1}).  

For our purposes, monoidal closure, in either its typed or untyped form,  is too strong: it describes situations where the computation is carried out in a single homogeneous system. Further, we do not expect, or require, universal computation from our heterotic systems. Instead, we take the notion of an adjunction between two functors as primitive, and expect to recover more familiar models of computation in the special case where the interacting systems are identical. 

The notion of an adjunction is simply a categorification of the concept of a Galois connection, thus two functors $\Gamma : \mathcal{C}_A \rightarrow \mathcal{C}_B$ and $\Delta : 
\mathcal{C}_B\rightarrow \mathcal{C}_A$ form an adjoint pair when $\mathcal{C}_A(\Gamma (X),Y) \ \cong \ \mathcal{C}%
_B(X,\Delta(Y)) $, for all $X\in Ob(\mathcal{C}_A)$, $Y\in Ob(\mathcal{C}_B)$.  
The duality provided by such an adjunction  allows us to 
model the mutual update
of system $A$ by system $B$ and system $B$ by system $A$, without requiring that system $B$ is fully  able to simulate the behaviour of system $A$, or vice versa. We are thus able to capture the sometimes
hidden symmetries we expect to find within such interactions. 

For concrete examples, we expect much more categorical structure; we are not claiming that the theory of adjunctions in itself will provide enough structure to give categorical semantics of heterotic systems. However, we take the existence of a suitable adjunction, between categories modelling dissimilar systems, as the basic defining characteristic of a heterotic system. Each concrete example will depend on the specific details of the two interacting systems. An illustrative example is available in
the categorical semantics approach of 
Abramsky and Coecke
\cite{AC04}, where
an adjunction (via its characterisation as unit/co-unit maps in a 2-category setting)
is used to describe creation of quantum systems from classical data,
and measurement of quantum systems (resulting in classical information). 

Thus, it 
appears that relatively simple category theory provides ready-made abstract conditions suitable for describing the mutual update of
distinct systems in heterotic computing, along with real concrete examples of how this works in certain settings.

\section{A heterotic refinement framework}
\label{sec:ref}

Given some suitable semantic framework,
such as the one outlined above,
it is necessary to cast it in a form suitable for
enabling the working programmer to analyse and develop novel heterotic
systems in (relatively) familiar ways. 
We suggest that a classical refinement framework is more appropriate than, 
say, a process algebra
approach, since this is more accessible and familiar to the working programmer. 

\begin{figure}[tp] 
\centering
(a) \scalebox{0.7}{\includegraphics{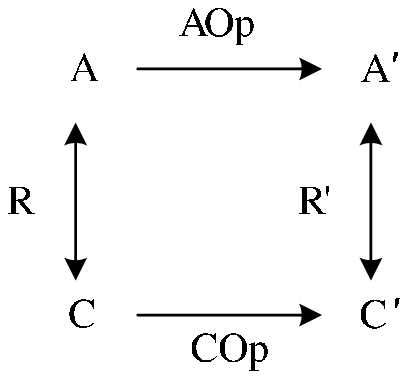}}\hspace{1cm}
(b) \scalebox{0.7}{\includegraphics{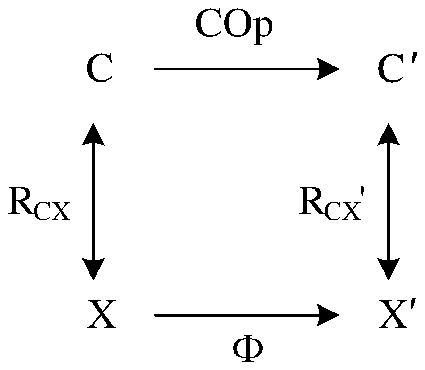}}
\caption{%
(a) A simulation, used to prove refinement;
(b) Physical and computational layer relationship
}
\label{fig:sim}
\end{figure}

State-and-operation refinement is the classical computational approach to program development.
It takes an abstract, possibly non-determin\-istic, specification 
of a state $A$ evolving under a sequence of operations $AOp$,
and \emph{refines} it (reducing non-determinism, changing data types) 
into a more concrete implementation 
with state $C$ and operations $COp$,
with the abstract state $A$ \emph{retrieved} from the concrete state $C$ 
through the retrieve relation $R$ (figure~\ref{fig:sim}a).
We have the refinement correctness requirement (ignoring non-determinism here for simplicity) that the diagram \emph{commute}
(we get the same value for $C'$ either way round):
\begin{equation}
R'(AOp(A)) = COp(R(A)) 
\end{equation} 

Usually the process of refinement stops
at a computational level suitably concrete to allow implementation,
such as a mid-level programming language.  
It can in principle be carried further.
Here we need to consider it all the way down to the physical implementation,
since we are interested in non-classical execution models.
So we continue refining from $C$ down to the physical level,
with a state $X$, that evolves under the laws of physics, $\Phi$.
The physical state variables in $X$
are again retrieved through  relation $R_{CX}$
as computational state variables in $C$
(figure~\ref{fig:sim}b).
Refinement reduces non-determinism
until we reach a completely deterministic implementation.
We classically think of the resolution of non-determinism being
under the control of the programmer, but when we reach the
physical layer we may be left with intrinsic non-determinism.
In the case of quantum computation,
while unitary quantum evolution is deterministic,
measurement of  quantum systems in general is not.
The programmer can either arrange for the algorithm to present
a final state with a deterministic measurement outcome (derandomization), or, 
accept that the computation may need to be repeated to increase the
probability of obtaining the required outcome.
%
%
Note that the induced computation $COp$ depends on both the physical system $\Phi$ and the viewing interpretation $R_{CX}$.
We would like this diagram to commute
(to get the same value for $X'$ either way round),
but there will be errors (measurement, noise)\footnote{%
Classical digital hardware is extremely engineered to ensure 
an exact boolean implementation;
this exactness cannot necessarily be assumed in the more general case.
}.
So we can at best require the inexact commutation
\begin{equation}
{R_{CX}}' (COp(C)) = \Phi(R_{CX}(C)) \pm \epsilon
\end{equation}

Retrenchment \cite{SS-WSCS04,SS-FI-06,Banach98,Banach2007301} is a form of inexact refinement.
It allows deviations from exact refinements by use of various forms of \emph{concedes} clauses;
analysis of the retrenchment concessions provides insight
into the way an implementation deviates from a pure refinement.
In particular, retrenchment has been applied to developing
discrete implementations of real number specifications \cite{Banach98},
and to finite implementations of unbounded natural number specifications,
which are necessarily inexact.
Also, it has been suggested as a laboratory for analysing and understanding emergent behaviour of complex systems \cite{SS-WSCS04}.

Retrenchment has its critics 
in the purist refinement community,
but we have argued elsewhere \cite{SS-FI-06} that these criticisms are invalid
in the context of real world engineering developments, even in the classical computing model.
Here we claim that (some suitably posed form of) retrenchment
is appropriate for casting non-exact computations in unconventional substrates in 
a refinement-like framework.
It would be used to analyse the size, nature, and
propagation of errors.

The usual classical refinement correctness rules allow
inputs to and outputs from the operations,
but require these to be the same at the abstract and concrete levels.
In previous work \cite{SS-YCS-347}, we have generalised these rules to 
allow refinement of i/o, too.
This necessitated the introduction of a \emph{finalisation} step,
that can be interpreted as the definition of the observation made on the system.
There is an \emph{initialisation} step, 
that we have  extended to interpret inputs analogously.
The finalisation of the most abstract level is usually the identity 
(we see the ``naked'' abstract i/o);
more concrete implementations have more sophisticated finalisations 
(eg, we see a bit stream, but view it, finalise it, as an integer) \cite{SS-REFINE05a}.
%
%
The correctness rule (again, ignoring non-determinism) is
\begin{equation}
AFin(A) = CFin(R(A)) 
\end{equation} 
This work has also been extended to the retrenchment arena.

A form of i/o refinement is necessary
to move between physical i/o variables and computational i/o variables.
For example, in the case of the NMR adder \cite{RBSS10}:
the physical level is the NMR;
the computational level is the NAND gate;
the initialisation is interpreting a frequency and a phase delay as a bit;
the finalisation is observing an integrated signal as a bit.
For this form of initialisation/finalisation to work in the analysis, it has
to be possible \emph{in principle} to provide all the inputs at the start of the computation,
and to observe (a record of) all the outputs at the end. 
This cannot be done for the individual layers of the heterotic computation, where the output
from one layer becomes the input to the other (it is closer to a Wegner interaction machine
architecture \cite{Wegner})
but can for the overall computation,
so we need to be careful about how we set up the analysis, and precisely
what we define as i/o. This step is crucial in our heterotic framework,
since, as stated earlier, the encoding and decoding processes 
(formalised as initialisation and finalisation) are
non-trivial in general.

We have an additional step in the NMR example \cite{RBSS10}, where the physical inputs and outputs are of different types, but the output from one step becomes the input to the next.
We perform a \emph{signal transduction} step here (integrals over Fourier transforms transduced to phases, that preserves the initialisation/finalisation interpretations).
This does not have an analogue in the refinement scenario,
because that does not include any link between the outputs of one operation and the inputs of the next.
This is important in the context of heterotic computing,
as there is potentially significant computation applied to outputs to produce the next inputs.
This computation is performed by the other part of the computer.

The base and controller levels can be implemented (refined) separately.
For example, in the quantum cluster state and the classical controller
(figure \ref{fig:cluster}a),
the state is set up initially, and the only operation performed 
in the base layer is measurement;
which measurement to perform is determined in the classical controller level
based on previous measurement results.
The measurement itself changes the state, which is part of the computation.
In NMR (figure \ref{fig:cluster}b), where the base level is the NMR gates;
the controller level is mere signal transduction -- this shows
that there is no sharp separation between the i/o refinement
and the computation (in this case it can be done in either).

These concrete models can be used as the basis for developing a
suitable form of refinement calculus. Possibly the closest pre-existing work
relating to this is the use of weakest precondition semantics to study
Grover's algorithm developed by d'Hondt and Panagaden
\cite{DHP06}
--- in particular, the way that a hybrid quantum/probabilistic setting is
modelled by the density matrix formalism. This gives a specific case
of the type of underlying logical rules that need to be preserved by the
refinement calculus, by analogy with the way that traditional program
refinement preserves the Hoare logic. However, in each concrete setting, 
the behaviour/logic preserved by the refinement process will be
different, and the formal calculus produced in each case will be heavily
dependent on the underlying categorical models.
Moreover, for non-discretised systems,
this relevant refinement calculus would need to be extended to a retrenchment
approach  to allow a well-defined and principled form of inexact refinement.
This would include analysis of propagation of errors \cite{Blakey2010}
(due to noise, and to drift), and techniques for correction and
control of these errors.

\section{Discussion and conclusions}
\label{sec:conc}

We have described a novel computational framework, heterotic computation, 
that can be used to combine computational systems from
different implementation paradigms in a principled and controlled manner,
to produce a computational system qualitatively different from either
in isolation.
We have outlined a semantic and refinement framework that could be used
to support such an approach.

One goal of such a framework is to analysise the efficiency
of a computational system.  Here we take a broad view of ``efficiency'':
it covers both the traditional scaling and complexity classes,
and also covers issues of real-time performance on real world scale problems.
Both views are important, and they do not necessarily coincide,
especially in combinations of disparate physical systems each being
exploited for its own particular computational capabilities.
As an example, the quantum community is developing ``hybrid computing''
\cite{spiller05b,spiller05a,milburn2000}, to create practical quantum
systems that can compute something non-trivial before errors come to dominate.
There efficiency gains from the theoretical complexity point of view
are considered later, only once the abstract theory is tackled
(for example, MBQC vs ancilla-driven quantum computation).
The heterotic framework, in both its categorical semantics and its
refinement/retrenchment calculus, allows for a range of efficiency
considerations, because it allows analysis of the computational
processes and error propagation in all the relevant parts of the system:
the individual layers, their interactions, and the overall system.
From this, both the complexity theoretic efficiency and the practical 
efficiency can be derived.


This is only the first step in such heterotic computation.
We have mentioned several areas that would need enhancement to
the simple framework we have started with:
where the base layer continues its computation whilst the controlling
layer is working, and where there is more than one layer.
A range of dynamical systems will contain continuously evolving layers;
one of the things the controlling layer will need to decide is when
to probe/perturb the base layer, to exploit its dynamics.
Additionally, further forms of parallelism
also need to be added to the framework.

We believe the heterotic approach is needed to ensure that the many
forms of unconventional computation can be exploited fully.
Each individual paradigm no longer need be distorted to achieve
Turing-completeness.
Instead, different components can be combined to form a more
powerful system, with each component doing what it does naturally,
and best.

\subsubsection*{Acknowledgments:}
VK is funded by a UK Royal Society University Research Fellowship.
We thank Matthias Bechmann, Rob Wagner, and Katherine Brown for
useful discussions.

\bibliography{publist}

\begin{thebibliography}{10}
\providecommand{\bibitemdeclare}[2]{}
\providecommand{\urlprefix}{Available at }
\providecommand{\url}[1]{\texttt{#1}}
\providecommand{\href}[2]{\texttt{#2}}
\providecommand{\urlalt}[2]{\href{#1}{#2}}
\providecommand{\doi}[1]{doi:\urlalt{http://dx.doi.org/#1}{#1}}
\providecommand{\bibinfo}[2]{#2}

\bibitemdeclare{article}{AC04}
\bibitem{AC04}
\bibinfo{author}{S.~Abramsky} \& \bibinfo{author}{B.~Coecke}
  (\bibinfo{year}{2004}): \emph{\bibinfo{title}{A categorical semantics of
  quantum protocols}}.
\newblock {\sl \bibinfo{journal}{Proc. IEEE Symp. Logic In Comp. Sci.}} , pp.
  \bibinfo{pages}{415--425}, \doi{10.1109/LICS.2004.1319636}.

\bibitemdeclare{article}{Adamatzky_2007}
\bibitem{Adamatzky_2007}
\bibinfo{author}{Andrew Adamatzky} (\bibinfo{year}{2007}):
  \emph{\bibinfo{title}{Physarum machines: encapsulating reaction-diffusion to
  compute spanning tree}}.
\newblock {\sl \bibinfo{journal}{Naturwissenschaften}}
  \bibinfo{volume}{94}(\bibinfo{number}{12}), pp. \bibinfo{pages}{975--980},
  \doi{10.1007/s00114-007-0276-5}.

\bibitemdeclare{book}{Amos}
\bibitem{Amos}
\bibinfo{author}{Martyn Amos} (\bibinfo{year}{2005}):
  \emph{\bibinfo{title}{Theoretical and Experimental DNA Computation}}.
\newblock \bibinfo{publisher}{Springer}, ISBN \bibinfo{isbn}{978-3-642-08504-8}.

\bibitemdeclare{article}{AB09}
\bibitem{AB09}
\bibinfo{author}{Janet Anders} \& \bibinfo{author}{Dan Browne}
  (\bibinfo{year}{2009}): \emph{\bibinfo{title}{Computational power of
  correlations}}.
\newblock {\sl \bibinfo{journal}{Phys.~Rev.~Lett.}} \bibinfo{volume}{102}, p.
  \bibinfo{pages}{050502}, \doi{10.1103/PhysRevLett.102.050502}.

\bibitemdeclare{article}{AOKBA10}
\bibitem{AOKBA10}
\bibinfo{author}{Janet Anders}, \bibinfo{author}{Daniel K.~L. Oi},
  \bibinfo{author}{Elham Kashefi}, \bibinfo{author}{Dan~E. Browne} \&
  \bibinfo{author}{Erika Andersson} (\bibinfo{year}{2010}):
  \emph{\bibinfo{title}{Ancilla-driven universal quantum computation}}.
\newblock {\sl \bibinfo{journal}{Phys.~Rev.~A}}
  \bibinfo{volume}{82}(\bibinfo{number}{2}), p. \bibinfo{pages}{020301},
  \doi{10.1103/PhysRevA.82.020301}.

\bibitemdeclare{inproceedings}{SS-WSCS04}
\bibitem{SS-WSCS04}
\bibinfo{author}{Richard Banach}, \bibinfo{author}{Czeslaw Jeske},
  \bibinfo{author}{Simon Fraser}, \bibinfo{author}{Richard Cross},
  \bibinfo{author}{Michael Poppleton}, \bibinfo{author}{Susan Stepney} \&
  \bibinfo{author}{Steven King} (\bibinfo{year}{2004}):
  \emph{\bibinfo{title}{Approaching the Formal Design and Development of
  Complex Systems: The Retrenchment Position}}.
\newblock In: {\sl \bibinfo{booktitle}{WSCS, IEEE ICECCS'04}}.
\newblock
  \urlprefix\url{http://citeseerx.ist.psu.edu/viewdoc/summary?doi=10.1.1.129.9231}.

\bibitemdeclare{article}{SS-FI-06}
\bibitem{SS-FI-06}
\bibinfo{author}{Richard Banach}, \bibinfo{author}{Czeslaw Jeske},
  \bibinfo{author}{Mike Poppleton} \& \bibinfo{author}{Susan Stepney}
  (\bibinfo{year}{2007}): \emph{\bibinfo{title}{Retrenching the Purse}}.
\newblock {\sl \bibinfo{journal}{Fundamenta Informaticae}}
  \bibinfo{volume}{77}, pp. \bibinfo{pages}{29--69}.
\newblock
  \urlprefix\url{http://citeseer.ist.psu.edu/viewdoc/summary?doi=10.1.1.61.8631}.

\bibitemdeclare{inproceedings}{Banach98}
\bibitem{Banach98}
\bibinfo{author}{Richard Banach} \& \bibinfo{author}{Mike Poppleton}
  (\bibinfo{year}{1998}): \emph{\bibinfo{title}{Retrenchment: an engineering
  variation on refinement}}.
\newblock In: {\sl \bibinfo{booktitle}{2nd Intl. B Conference}}, {\sl
  \bibinfo{series}{LNCS}} \bibinfo{volume}{1393},
  \bibinfo{publisher}{Springer}, pp. \bibinfo{pages}{129--147},
  \doi{10.1007/BFb0053358}.

\bibitemdeclare{article}{Banach2007301}
\bibitem{Banach2007301}
\bibinfo{author}{Richard Banach}, \bibinfo{author}{Mike Poppleton},
  \bibinfo{author}{Czeslaw Jeske} \& \bibinfo{author}{Susan Stepney}
  (\bibinfo{year}{2007}): \emph{\bibinfo{title}{Engineering and theoretical
  underpinnings of retrenchment}}.
\newblock {\sl \bibinfo{journal}{Sci. Comp. Prog.}}
  \bibinfo{volume}{67}(\bibinfo{number}{2-3}), pp. \bibinfo{pages}{301--329},
  \doi{10.1016/j.scico.2007.04.002}.

\bibitemdeclare{article}{BSS11}
\bibitem{BSS11}
\bibinfo{author}{M.~Bechmann}, \bibinfo{author}{A.~Sebald} \&
  \bibinfo{author}{S.~Stepney} (\bibinfo{year}{2011}):
  \emph{\bibinfo{title}{Boolean logic-gate design principles in unconventional
  computers: an {NMR} case study}}.
\newblock {\sl \bibinfo{journal}{International J.~Unconventional Computing}}
  \bibinfo{note}{In press}.

\bibitemdeclare{inproceedings}{SS-ICES2010}
\bibitem{SS-ICES2010}
\bibinfo{author}{Matthias Bechmann}, \bibinfo{author}{Angelika Sebald} \&
  \bibinfo{author}{Susan Stepney} (\bibinfo{year}{2010}):
  \emph{\bibinfo{title}{From binary to continuous gates -- and back again}}.
\newblock In: {\sl \bibinfo{booktitle}{ICES 2010}}, {\sl
  \bibinfo{series}{LNCS}} \bibinfo{volume}{6274},
  \bibinfo{publisher}{Springer}, pp. \bibinfo{pages}{335--347},
  \doi{10.1007/978-3-642-15323-5_29}.

\bibitemdeclare{article}{Blakey2010}
\bibitem{Blakey2010}
\bibinfo{author}{Ed~Blakey} (\bibinfo{year}{2010}):
  \emph{\bibinfo{title}{Unconventional complexity measures for unconventional
  computers}}.
\newblock {\sl \bibinfo{journal}{Natural Computing}}
  \doi{10.1007/s11047-010-9226-9}.

\bibitemdeclare{misc}{BHKM11}
\bibitem{BHKM11}
\bibinfo{author}{K.~L. Brown}, \bibinfo{author}{C.~Horsman},
  \bibinfo{author}{V.~M. Kendon} \& \bibinfo{author}{W.~J. Munro}
  (\bibinfo{year}{2011}): \emph{\bibinfo{title}{Layer by layer generation of
  cluster states}}.
\newblock \urlprefix\url{http://arxiv.org/abs/1111.1774v1}.

\bibitemdeclare{article}{BDKM10}
\bibitem{BDKM10}
\bibinfo{author}{Katherine~L. Brown}, \bibinfo{author}{Suvabrata De},
  \bibinfo{author}{Viv Kendon} \& \bibinfo{author}{William~J. Munro}
  (\bibinfo{year}{2011}): \emph{\bibinfo{title}{Ancilla-based quantum
  simulation}}.
\newblock {\sl \bibinfo{journal}{New J. Phys.}} \bibinfo{volume}{13}, p.
  \bibinfo{pages}{095007}, \doi{10.1088/1367-2630/13/9/095007}.

\bibitemdeclare{article}{SS-REFINE05a}
\bibitem{SS-REFINE05a}
\bibinfo{author}{John~A. Clark}, \bibinfo{author}{Susan Stepney} \&
  \bibinfo{author}{Howard Chivers} (\bibinfo{year}{2005}):
  \emph{\bibinfo{title}{Breaking the Model: finalisation and a taxonomy of
  security attacks}}.
\newblock {\sl \bibinfo{journal}{ENTCS}}
  \bibinfo{volume}{137}(\bibinfo{number}{2}), pp. \bibinfo{pages}{225--242},
  \doi{10.1016/j.entcs.2005.04.033}.

\bibitemdeclare{techreport}{SS-YCS-347}
\bibitem{SS-YCS-347}
\bibinfo{author}{D.~Cooper}, \bibinfo{author}{S.~Stepney} \&
  \bibinfo{author}{J.~Woodcock} (\bibinfo{year}{2002}):
  \emph{\bibinfo{title}{Derivation of {Z} Refinement Proof Rules: forwards and
  backwards rules incorporating input/output refinement}}.
\newblock \bibinfo{type}{Technical Report} \bibinfo{number}{YCS-2002-347},
  \bibinfo{institution}{Department of Computer Science, University of York}.
\newblock
  \urlprefix\url{http://www.cs.york.ac.uk/ftpdir/reports/2002/YCS/347/YCS-2002-347.pdf}.

\bibitemdeclare{article}{Cory2000}
\bibitem{Cory2000}
\bibinfo{author}{David~G. Cory} et~al. (\bibinfo{year}{2000}):
  \emph{\bibinfo{title}{{NMR} Based Quantum Information Processing:
  Achievements and Prospects}}.
\newblock {\sl \bibinfo{journal}{Fortschritte der Physik}}
  \bibinfo{volume}{48}(\bibinfo{number}{9--11}), pp. \bibinfo{pages}{875--907},
  \doi{10.1002/1521-3978(200009)48:9/11}.

\bibitemdeclare{article}{DHP06}
\bibitem{DHP06}
\bibinfo{author}{E.~{d'H}ondt} \& \bibinfo{author}{P.~Panangaden}
  (\bibinfo{year}{2006}): \emph{\bibinfo{title}{Quantum weakest
  preconditions}}.
\newblock {\sl \bibinfo{journal}{Math. Struct. Comp. Sci.}}
  \bibinfo{volume}{16}(\bibinfo{number}{3}), pp. \bibinfo{pages}{429--451},
  \doi{10.1017/S0960129506005251}.

\bibitemdeclare{article}{DiV00}
\bibitem{DiV00}
\bibinfo{author}{David~P. DiVincenzo} (\bibinfo{year}{2000}):
  \emph{\bibinfo{title}{The Physical Implementation of Quantum Computation}}.
\newblock {\sl \bibinfo{journal}{Fortschritte der Physik}}
  \bibinfo{volume}{48}(\bibinfo{number}{9--11}), pp. \bibinfo{pages}{771--783},
  \doi{10.1002/1521-3978(200009)48:9/11}.
\newblock \bibinfo{note}{ArXiv:quant-ph/0002077v3}.

\bibitemdeclare{article}{PH1}
\bibitem{PH1}
\bibinfo{author}{Peter Hines} (\bibinfo{year}{1999}): \emph{\bibinfo{title}{The
  categorical theory of self-similarity}}.
\newblock {\sl \bibinfo{journal}{Theory and Applications of Categories}}
  \bibinfo{volume}{6}, pp. \bibinfo{pages}{33--46}.
\newblock \urlprefix\url{http://emis.math.ca/journals/TAC/volumes/6/n3/n3.pdf}.

\bibitemdeclare{article}{PH2}
\bibitem{PH2}
\bibinfo{author}{Peter Hines} (\bibinfo{year}{2003}): \emph{\bibinfo{title}{A
  categorical framework for finite state machines}}.
\newblock {\sl \bibinfo{journal}{Mathematical Structures in Computer Science}}
  \bibinfo{volume}{13}, pp. \bibinfo{pages}{451--480},
  \doi{10.1017/S0960129503003931}.

\bibitemdeclare{article}{HBMK11}
\bibitem{HBMK11}
\bibinfo{author}{Clare Horsman}, \bibinfo{author}{Katherine~L. Brown},
  \bibinfo{author}{William~J. Munro} \& \bibinfo{author}{Vivien~M. Kendon}
  (\bibinfo{year}{2011}): \emph{\bibinfo{title}{Reduce, reuse, recycle for
  robust cluster-state generation}}.
\newblock {\sl \bibinfo{journal}{Phys. Rev. A}}
  \bibinfo{volume}{83}(\bibinfo{number}{4}), p. \bibinfo{pages}{042327},
  \doi{10.1103/PhysRevA.83.042327}.

\bibitemdeclare{article}{Jones2011}
\bibitem{Jones2011}
\bibinfo{author}{Jonathan~A. Jones} (\bibinfo{year}{2011}):
  \emph{\bibinfo{title}{Quantum computing with {NMR}}}.
\newblock {\sl \bibinfo{journal}{Progress in Nuclear Magnetic Resonance
  Spectroscopy}} \bibinfo{volume}{59}, pp. \bibinfo{pages}{91--120},
  \doi{10.1016/j.pnmrs.2010.11.001}.

\bibitemdeclare{misc}{Jozsa05}
\bibitem{Jozsa05}
\bibinfo{author}{Richard Jozsa} (\bibinfo{year}{2005}):
  \emph{\bibinfo{title}{An introduction to measurement based quantum
  computation}}.
\newblock \urlprefix\url{http://arxiv.org/abs/quant-ph/0508124}.

\bibitemdeclare{inproceedings}{KSSBHW11}
\bibitem{KSSBHW11}
\bibinfo{author}{V.~Kendon}, \bibinfo{author}{A.~Sebald},
  \bibinfo{author}{S.~Stepney}, \bibinfo{author}{Matthias Bechmann},
  \bibinfo{author}{Peter Hines} \& \bibinfo{author}{Robert~C. Wagner}
  (\bibinfo{year}{2011}): \emph{\bibinfo{title}{Heterotic computing}}.
\newblock In: {\sl \bibinfo{booktitle}{Unconventional Computation, LNCS}},
  \bibinfo{volume}{6714}, \bibinfo{publisher}{Springer}, pp.
  \bibinfo{pages}{113--124}, \doi{10.1007/978-3-642-21341-0_16}.

\bibitemdeclare{article}{Kuhnert_1989}
\bibitem{Kuhnert_1989}
\bibinfo{author}{L.~Kuhnert}, \bibinfo{author}{K.~Agladze} \&
  \bibinfo{author}{V.~Krinsky} (\bibinfo{year}{1989}):
  \emph{\bibinfo{title}{Image processing using light-sensitive chemical
  waves}}.
\newblock {\sl \bibinfo{journal}{Nature}} \bibinfo{volume}{337}, pp.
  \bibinfo{pages}{244--247}, \doi{10.1038/337244a0}.

\bibitemdeclare{book}{LS}
\bibitem{LS}
\bibinfo{author}{J.~Lambek} \& \bibinfo{author}{P.~J. Scott}
  (\bibinfo{year}{1988}): \emph{\bibinfo{title}{An introduction to higher-order
  categorical logic}}.
\newblock {\sl \bibinfo{series}{Cambridge Studies in Advanced
  Mathematics}}~\bibinfo{volume}{7}, \bibinfo{publisher}{Cambridge University
  Press}, ISBN \bibinfo{isbn}{9780521356534}.

\bibitemdeclare{article}{lloyd99b}
\bibitem{lloyd99b}
\bibinfo{author}{Seth Lloyd} \& \bibinfo{author}{Samuel~L Braunstein}
  (\bibinfo{year}{1999}): \emph{\bibinfo{title}{Quantum computation over
  continuous variables}}.
\newblock {\sl \bibinfo{journal}{Phys.~Rev.~Lett.}} \bibinfo{volume}{82}, p.
  \bibinfo{pages}{1784}, \doi{10.1103/PhysRevLett.82.1784}.

\bibitemdeclare{book}{MCL}
\bibitem{MCL}
\bibinfo{author}{Saunders {Mac Lane}} (\bibinfo{year}{1971}):
  \emph{\bibinfo{title}{Categories for the working mathematician}}.
\newblock \bibinfo{series}{Graduate Texts in Mathematics, 1st Ed.},
  \bibinfo{publisher}{Springer Verlag}, ISBN \bibinfo{isbn}{0387900357}.

\bibitemdeclare{article}{milburn2000}
\bibitem{milburn2000}
\bibinfo{author}{G.~J. Milburn}, \bibinfo{author}{S.~Schneider} \&
  \bibinfo{author}{D.~F.~V. James} (\bibinfo{year}{2000}):
  \emph{\bibinfo{title}{Ion Trap Quantum Computing with Warm Ions}}.
\newblock {\sl \bibinfo{journal}{Fortschr.~Phys.}} \bibinfo{volume}{48}, pp.
  \bibinfo{pages}{801--810},
  \doi{10.1002/1521-3978(200009)48:9/11<801::AID-PROP801>3.0.CO;2-1}.

\bibitemdeclare{article}{mills08a}
\bibitem{mills08a}
\bibinfo{author}{Jonathan~W. Mills} (\bibinfo{year}{2008}):
  \emph{\bibinfo{title}{The nature of the Extended Analog Computer}}.
\newblock {\sl \bibinfo{journal}{Physica D: Nonlinear Phenomena}}
  \bibinfo{volume}{237}(\bibinfo{number}{9}), pp. \bibinfo{pages}{1235--1256},
  \doi{10.1016/j.physd.2008.03.041}.

\bibitemdeclare{article}{Motoike2005107}
\bibitem{Motoike2005107}
\bibinfo{author}{Ikuko~N. Motoike} \& \bibinfo{author}{Andrew Adamatzky}
  (\bibinfo{year}{2005}): \emph{\bibinfo{title}{Three-valued logic gates in
  reaction-diffusion excitable media}}.
\newblock {\sl \bibinfo{journal}{Chaos, Solitons \& Fractals}}
  \bibinfo{volume}{24}(\bibinfo{number}{1}), pp. \bibinfo{pages}{107--114},
  \doi{10.1016/S0960-0779(04)00461-8}.

\bibitemdeclare{article}{RB01}
\bibitem{RB01}
\bibinfo{author}{Robert Raussendorf} \& \bibinfo{author}{Hans~J Briegel}
  (\bibinfo{year}{2001}): \emph{\bibinfo{title}{A One-Way Quantum Computer}}.
\newblock {\sl \bibinfo{journal}{Phys.~Rev.~Lett.}} \bibinfo{volume}{86}, pp.
  \bibinfo{pages}{5188--5191}, \doi{10.1103/PhysRevLett.86.5188}.

\bibitemdeclare{article}{RBSS10}
\bibitem{RBSS10}
\bibinfo{author}{M.~Rosell\'{o}-Merino}, \bibinfo{author}{M.~Bechmann},
  \bibinfo{author}{A.~Sebald} \& \bibinfo{author}{S.~Stepney}
  (\bibinfo{year}{2010}): \emph{\bibinfo{title}{Classical computing in nuclear
  magnetic resonance.}}
\newblock {\sl \bibinfo{journal}{International J.~Unconventional Computing}}
  \bibinfo{volume}{6}(\bibinfo{number}{3--4}), pp. \bibinfo{pages}{163--195}.
\newblock
  \urlprefix\url{http://www-users.cs.york.ac.uk/susan/bib/ss/nonstd/ijnmc09.pdf}.

\bibitemdeclare{article}{silvagraca04a}
\bibitem{silvagraca04a}
\bibinfo{author}{D.~{Silva Gra\c{c}a}} (\bibinfo{year}{2004}):
  \emph{\bibinfo{title}{Some Recent Developments on {S}hannon's {GPAC}}}.
\newblock {\sl \bibinfo{journal}{Math.~Log.~Quart.}}
  \bibinfo{volume}{50}(\bibinfo{number}{4--5}), pp. \bibinfo{pages}{473--485},
  \doi{10.1002/malq.200310113}.

\bibitemdeclare{article}{spiller05a}
\bibitem{spiller05a}
\bibinfo{author}{T.~P. Spiller}, \bibinfo{author}{W.~J. Munro},
  \bibinfo{author}{S.~D. Barrett} \& \bibinfo{author}{P.~Kok}
  (\bibinfo{year}{2005}): \emph{\bibinfo{title}{An introduction to quantum
  information processing: applications and realisations}}.
\newblock {\sl \bibinfo{journal}{Comptemporary Physics}} \bibinfo{volume}{46},
  p. \bibinfo{pages}{407}, \doi{10.1080/00107510500293261}.

\bibitemdeclare{article}{spiller05b}
\bibitem{spiller05b}
\bibinfo{author}{T.~P. Spiller}, \bibinfo{author}{Kae Nemoto},
  \bibinfo{author}{Samuel~L. Braunstein}, \bibinfo{author}{W.~J. Munro},
  \bibinfo{author}{P.~van Loock} \& \bibinfo{author}{G.~J. Milburn}
  (\bibinfo{year}{2006}): \emph{\bibinfo{title}{Quantum Computation by
  Communication}}.
\newblock {\sl \bibinfo{journal}{New J.~Phys.}} \bibinfo{volume}{8},
  p.~\bibinfo{pages}{30}, \doi{10.1088/1367-2630/8/2/030}.

\bibitemdeclare{article}{SS-PhysicaD-08}
\bibitem{SS-PhysicaD-08}
\bibinfo{author}{Susan Stepney} (\bibinfo{year}{2008}):
  \emph{\bibinfo{title}{The Neglected Pillar of Material Computation}}.
\newblock {\sl \bibinfo{journal}{Physica D: Nonlinear Phenomena}}
  \bibinfo{volume}{237}(\bibinfo{number}{9}), pp. \bibinfo{pages}{1157--1164},
  \doi{10.1016/j.physd.2008.01.028}.

\bibitemdeclare{article}{Toth_1995}
\bibitem{Toth_1995}
\bibinfo{author}{{\'A}gota T{\'o}th} \& \bibinfo{author}{Kenneth Showalter}
  (\bibinfo{year}{1995}): \emph{\bibinfo{title}{Logic gates in excitable
  media}}.
\newblock {\sl \bibinfo{journal}{J. Chem. Phys}} \bibinfo{volume}{103}, pp.
  \bibinfo{pages}{2058--2066}, \doi{10.1063/1.469732}.

\bibitemdeclare{article}{Tucker2010}
\bibitem{Tucker2010}
\bibinfo{author}{Rodney~S. Tucker} (\bibinfo{year}{2010}):
  \emph{\bibinfo{title}{The role of optics in computing}}.
\newblock {\sl \bibinfo{journal}{Nature Photonics}} \bibinfo{volume}{4}, p.
  \bibinfo{pages}{405}, \doi{10.1038/nphoton.2010.162}.

\bibitemdeclare{article}{Wegner}
\bibitem{Wegner}
\bibinfo{author}{Peter Wegner} (\bibinfo{year}{1997}):
  \emph{\bibinfo{title}{Why interaction is more powerful than algorithms}}.
\newblock {\sl \bibinfo{journal}{CACM}} \bibinfo{volume}{40}, pp.
  \bibinfo{pages}{80--91}, \doi{10.1145/253769.253801}.

\bibitemdeclare{incollection}{Woods2008}
\bibitem{Woods2008}
\bibinfo{author}{Damien Woods} \& \bibinfo{author}{Thomas~J. Naughton}
  (\bibinfo{year}{2008}): \emph{\bibinfo{title}{Parallel and Sequential Optical
  Computing}}.
\newblock In: {\sl \bibinfo{booktitle}{Optical SuperComputing}}, {\sl
  \bibinfo{series}{LNCS}} \bibinfo{volume}{5172},
  \bibinfo{publisher}{Springer}, pp. \bibinfo{pages}{70--86},
  \doi{10.1007/978-3-540-85673-3_6}.

\end{thebibliography}

\end{document}